\documentclass[twocolumn,twoside,slac_two]{revtex4}
\usepackage{graphicx}
\usepackage{fancyhdr}
\usepackage{subfig}
\begin{document}

\title{Status of EXO-200}

%

\author{Nicole Ackerman on behalf of the EXO Collaboration}
\affiliation{SLAC National Accelerator Laboratory, Stanford CA USA}

\begin{abstract}
EXO-200 is the first phase of the Enriched Xenon Observatory (EXO) experiment, which searches for neutrinoless double beta decay in $^{136}$Xe to measure the mass and probe the Majorana nature of the neutrino.  EXO-200 consists of 200 kg of liquid Xe enriched to 80\% in $^{136}$Xe in an ultra-low background TPC.  Energy resolution is enhanced through the simultaneous collection of scintillation light using Large Area Avalanche Photodiodes (LAAPD's) and ionization charge. It is being installed at the WIPP site in New Mexico, which provides a 2000 meter water-equivalent overburden.  EXO-200 will begin taking data in 2009, with the expected two-year sensitivity to the half-life for neutrinoless double beta decay of $6.4 \times 10^{25}$ years. According to the most recent nuclear matrix element calculations, this corresponds to an effective Majorana neutrino mass of 0.13 to 0.19 eV. It will also measure the two neutrino mode for the first time in $^{136}$Xe.  
\end{abstract}

\maketitle

\thispagestyle{fancy}


\section{Introduction}

While reactors and accelerators can answer many questions about the neutrino through oscillation measurements, many remain that can be elegantly addressed through double beta decay experiments \cite{avignone_double_2008}.  It is still unknown whether neutrinos have a majorana nature or what the absolute mass scale of the neutrino is.  EXO-200 is one of many first-generation experiments searching for neutrinoless double beta decay and working towards ton-scale experiments with very promising discovery potential.

\section{Double Beta Decay}

\begin{figure}[h]
\centering
\subfloat[Standard Model \label{fig:bb2n}]{\includegraphics[width=35mm]{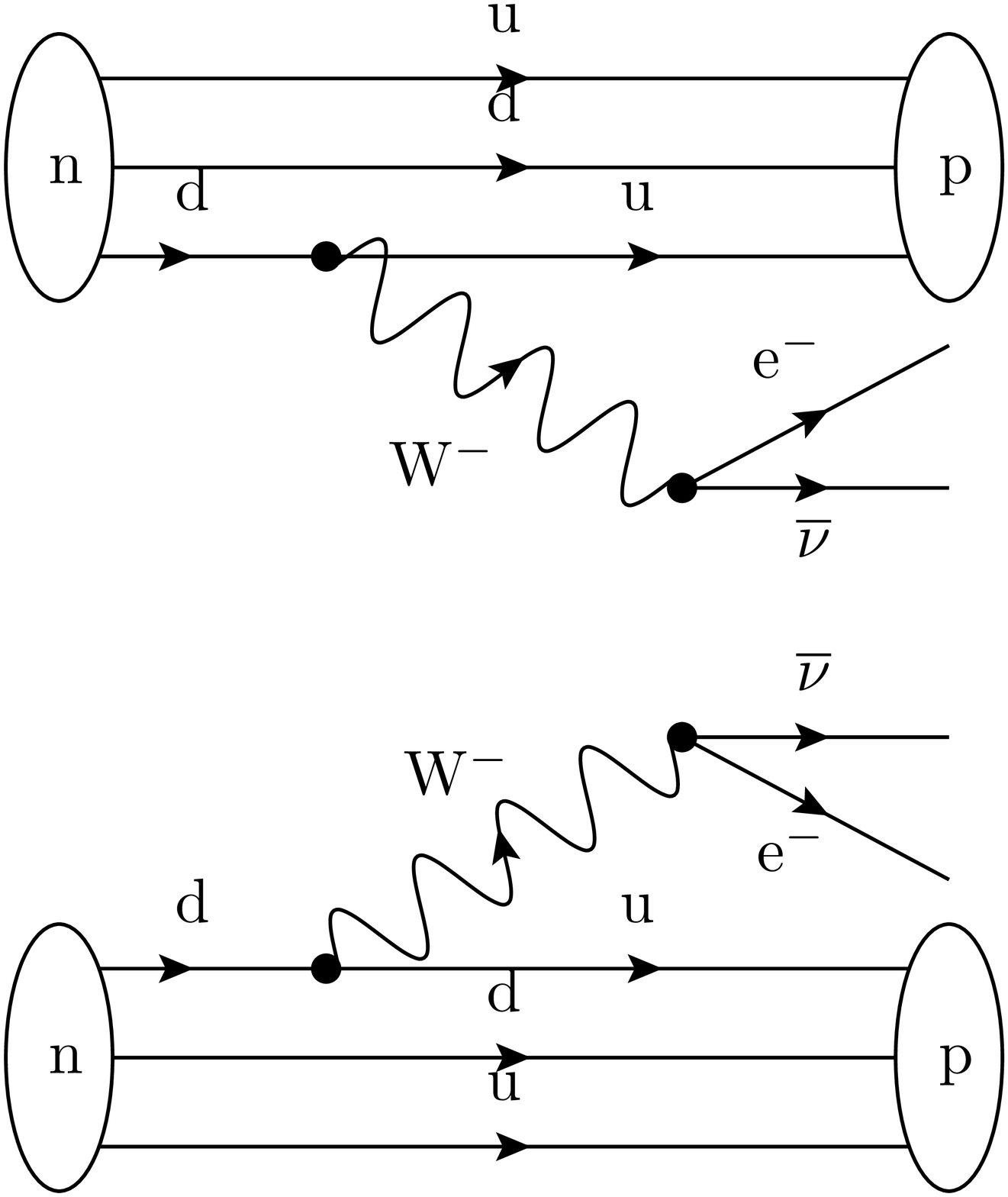}}
\hspace{6mm}
\subfloat[Neutrinoless \label{fig:bb0n}]{\includegraphics[width=35mm]{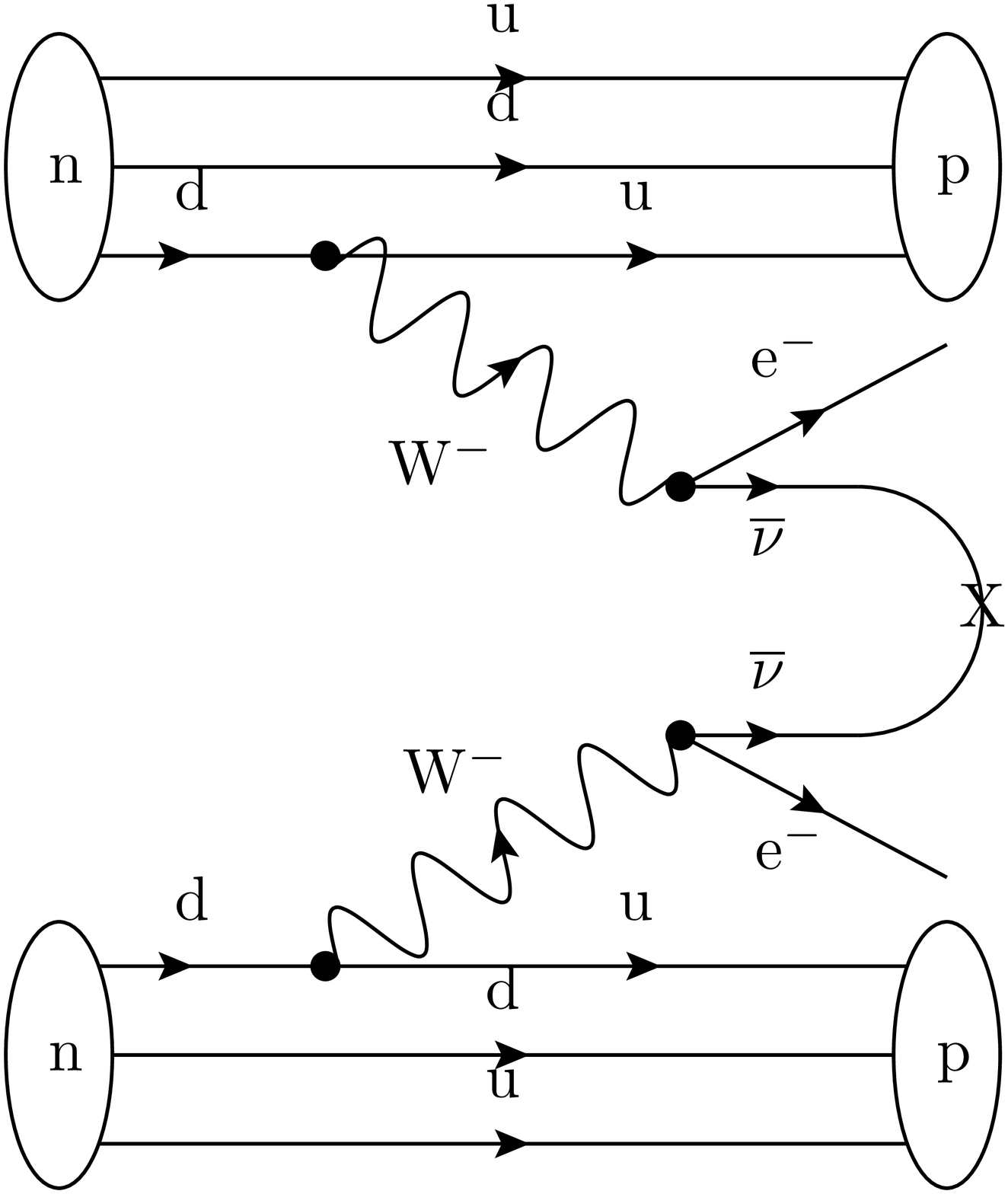}}
\caption{Feynman diagrams for the two double beta decay processes, showing that they have the same nuclear interaction but different final states.} \label{fig:bb}
\end{figure}

Single beta decay is a well-understood nuclear process by which a neutron decays to a proton, emitting an electron and an anti-electron neutrino in the process.  Double beta decay is a second-order standard model process where this occurs twice in the same nucleus, shown in Fig. \ref{fig:bb2n}.  The rate is exceedingly small, so only nuclei which are energetically forbidden from undergoing single beta decay are promising candidates for observing the double beta decay signal.  Two neutrino double beta decay ($2\nu \beta \beta$) has been observed in many isotopes, though not yet in $^{136}$Xe.
Neutrinoless double beta decay ($0\nu \beta \beta$) is forbidden by the standard model, shown in Fig. \ref{fig:bb0n}.  This process can only occur if the neutrino is its own anti-particle, so observation of $0\nu \beta \beta$ indicates that the neutrino is a Majorana particle.  While the half-life of $2\nu \beta \beta$ ($T_{1/2}^{2\nu }$) depends on phase space and nuclear matrix elements (NME), the $0\nu \beta \beta$ half-life ($T_{1/2}^{0\nu }$) is also dependent upon the effective mass of the neutrino.  Hence, a measurement of the rate of $0\nu \beta \beta$ provides a measurement of the neutrino mass scale.

The two processes can be clearly differentiated through their spectra.  $2\nu \beta \beta$ is a 4-body decay with energy lost to both neutrinos, while $0\nu \beta \beta$ is a two-body decay in which the sum of the energies carried by the electrons equals the Q-value of the decay.  The neutrinoless signal will appear as a peak, the width determined by detector resolution, at the endpoint of the $2 \nu \beta \beta$ spectrum.  Hence, it is essential to optimize resolution and eliminate backgrounds in that region.

\begin{figure*}[ht]
\centering
\subfloat[One half of the detector, as viewed from the cathode plane. The aluminum stand is only for construction and is not part of the detector.\label{fig:detector}]{\includegraphics[width=80mm]{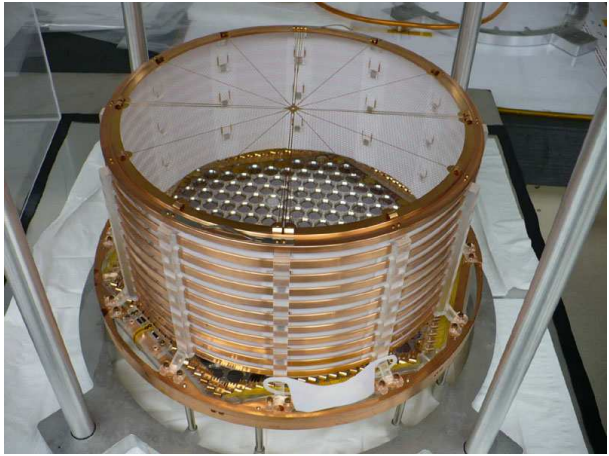}}
\hspace{5mm}
\subfloat[The TPC inserted into the vessel, with the attached cryostat door, as viewed from the bottom of the APD plane. The legs contain the readout flexcircuits and are the conduits for xenon circulation.\label{fig:vessel}]{\includegraphics[width=80mm]{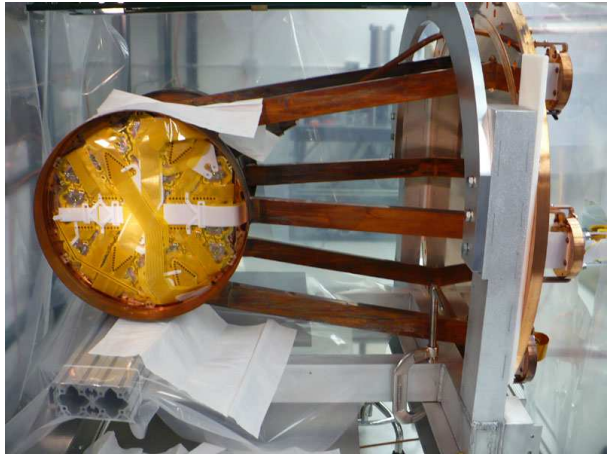}}
\caption{These images show the finished detector in the underground cleanrooms at Stanford University.} \label{fig:detector_const}
\end{figure*}

\section{Enriched Xenon Observatory}

The Enriched Xenon Observatory (EXO) Collaboration studies double beta decay in the isotope Xenon 136 \cite{avignone_double_2008}\cite{akimov_exo:advanced_2005}.  $^{136}$Xe is an ideal isotope for the study of $0\nu\beta\beta$.  Liquid xenon is a scintillator, and it has been shown that collecting both scintillation light and ionization signals significantly improves energy resolution \cite{conti_correlated_2003}\cite{aprile_observation_2007}.  Using a liquid or gas allows continuous circulation and purification of the xenon, further reducing contaminants.  A high Q-value places the region of interest for $0\nu\beta\beta$ above decay lines, so $^{136}$Xe is advantageous with Q=2.48 MeV \cite{redshaw_mass_2007}.  Additionally, xenon allows a unique technology of barium tagging.  Every double beta decay event will produce a Ba$^{++}$, while radioactive backgrounds will not.  By being able to identify single barium ions, all radioactive backgrounds can be excluded \cite{danilov_detection_2000}. While natural xenon has a relatively small abundance of $^{136}$Xe, it is relatively easy to enrich to 80\% using centrifuges, which also removes the lighter impurities.  
          
\section{200 kg Experiment}

One component of the EXO project is EXO-200, an experiment using 200 kg of liquid xenon enriched to 80\% in $^{136}$Xe.  EXO-200 is a low-background detector made of high radiopurity materials, built and operated in a clean room, and installed underground.  The detector is a time projection chamber (TPC) with collection of scintillation and ionization signals.  Construction of the detector is finished at Stanford and it will be soon installed at the Waste Isolation Pilot Plant (WIPP), where the infrastructure is undergoing final testing.  EXO-200 serves as a prototype for liquid xenon and low-background technologies. Its measurements will include $T_{1/2}^{2\nu }$ and $T_{1/2}^{0\nu}$ in $^{136}$Xe, with a competitive sensitivity to the effective neutrino mass,  $\langle m_{\beta\beta} \rangle$. 

\subsection{Design and Construction}

The TPC was designed to maximize fiducial volume, maximize resolution, and minimize possible contaminants.  Radiopurity is the most constraining requirment, prohibiting normal wiring, adhesives, and many common materials.  All of the components are spring mounted with phosphorbronze spring and screws, which are platinum plated at electrical contacts.  Figure \ref{fig:detector} shows one half of the detector where the acrylic and high purity-copper frame components are visible.  The read out cables are specially-made flexcircuits of copper on kapton, as can be seen on the end of the detector in Fig. \ref{fig:vessel}.    

The TPC is cylindrical, with the cathode plane in the center and detection planes on both ends.  The cathode plane and detection wires are photoetched phosphorbronze, chosen from the vendor that could demonstrate the best radiopurity.  There are two wire planes on each end, one being held at -4 kV and the second at ground.  The cathode was designed to run at up to -75 kV but may be operated at a lower voltage if it provides better energy resolution. The first wire plane provides an inductive signal while the second plane collects the electrons.  The two planes are rotated by $60^\circ$ with respect to each other, so the two signals provide an x-y coordinate for the event.  Along the length of the cylinder are copper field shaping rings, separated by high radiopurity resistors of 640M$\Omega$. 

The scintillation light is collected at the ends of the TPC by 234 Large Area Avalanche Photodiodes (LAAPDs), operated at a gain of $\sim$100 and a QE of $\sim$1.  The LAAPDs are mounted on a low-activity copper platter on both ends of the TPC in a triangular pattern.  The side facing the bulk of the xenon is plated with aluminum and magnesium fluoride for increased photon collection while the contact side is plated with gold for better contact to the LAAPDs.  The LAAPDs are ganged in nominal groups of 7 for 37 readout channels per side.  To increase photon collection, the inside of the detector is lined with teflon sheets which provides a high reflectivity of the $\sim$178 nm scintillation light.  See \cite{neilson_characterization_2009} for further discussion of the LAAPDs.

The liquid xenon is kept at a pressure at $\sim$1.5 atmospheres and a temperature of $\sim$170 K in a cylindrical vessel.  The vessel is thin-walled (1.5 mm), 40 cm long and 40 cm in diameter, and made of low-activity copper.  The vessel is contained in a low-activity copper cryostat, filled with HFE-7000, a cryogenic buffer fluid providing a thermal bath to the vessel and an innermost $\gamma$ and neutron shield. There is a vacuum insulation layer, including thermal radiation-blocking super-insulation, between the inner cryostat and a second low-activity copper cryostat.  The outer cryostat is contained within a nitrogen-purged tent for radon-prevention and is surrounded by a lead shield. The experiment is operated inside a class-100 clean room at the WIPP site.  On the outside of the cleanrooms are scintillation counters providing a muon veto.

All low-activity copper has been stored and machined underground.  Care was taken to prevent cosmic activation of materials.  The detector itself was assembled in an underground clean room located at Stanford University.  The cryogenic systems were initially assembled and commissioned in 6 clean room modules located at Stanford University.  The clean rooms, including the cryostat and associated plumbing, were then shipped to WIPP where they were commissioned a second time.  Throughout the design, procurement, and assembly processes, measurements were taken of the U, Th, and K contamination in all materials which are characterized in a large database \cite{leonard_systematic_2008}.  Measurements indicate we will achieve our goal background level in the $0\nu$ region of interest.

\subsection{Waste Isolation Pilot Plant}

EXO-200 is installed underground at the Waste Isolation Pilot Plant near Carlsbad, New Mexico.  The WIPP site is a Department of Energy facility used for the permanent storage of nuclear waste.  However, the waste is not a problem as the waste has low level radioactivity and is stored on the other side of the mine from the North Experimental Area (NExA).  The salt at WIPP provides an excellent low-background environment.  The radioactive contaminants in the salt are less concentrated that in surface soil for uranium, thorium, and potassium \cite{esch_cosmic_2005} and the background neutron level is relatively low at 332 $\pm$ 148 neutrons m$^2$d$^1$ \cite{balbes_evaluation_1997}.  The experiment is installed at a depth of 2150 feet (655 meters), which provides an overburden of 2000 meters water equivalent.
  

\begin{table*}[ht]
\begin{center}
\caption{Predicted number of $2\nu\beta\beta$ events in EXO-200 in one year, based up previous experimental limits for  $T_{1/2}^{2\nu }(^{136}$Xe) and theoretical calculations.\label{tab:2nsens}}
\begin{tabular}{|l|c|c|}
\hline
 & $T_{1/2}$(yr) & events/year \\
\hline
Experimental Limit & & \\
\hline
Luscher et al \cite{luscher_search_1998} & $>3.6 \times 10^{20} $ & $<$ 1.3 M \\
Gavriljuk et al  \cite{gavriljuk_results_2005} & $>8.5 \times 10^{21}$ & $<$ 60 k \\
Bernabei et al  \cite{bernabei_investigation_2002} & $>1.0 \times 10^{22}$ & $<$ 48 k \\
\hline
Theoretical prediction ($T_{1/2}^{\text{max}}$) & & \\
\hline
QRPA (Staudt et al) \cite{staudt_calculation_1990} & $=2.1 \times 10^{22}$ & =23 k\\
QRPA (Engel et al) \cite{engel_nuclear_1988} & $= 8.2  \times 10^{20} $ & =0.62 M \\
NSM (Caurier et al) \cite{caurier_shell_1996} & $=2.0 \times 10^{21}$ & =0.25 M \\
\hline
\end{tabular}
\end{center}
\end{table*}

\subsection{Sensitivities}

EXO-200 will search for $0\nu \beta \beta$ and $2\nu \beta \beta$ in $^{136}$Xe.  While $2\nu \beta \beta$ has not yet been observed in $^{136}$Xe \cite{bernabei_investigation_2002}, the theoretical predictions (see Tab. \ref{tab:2nsens}) indicate a half-life that will provide a large signal, likely in the tens to hundreds of thousands of events a year. However, the signal for $0\nu \beta \beta$ may be many orders of magnitude below the signal for $2\nu \beta \beta$.  Table \ref{tab:0nsens200} shows the predicted level at which EXO-200 will set a limit for the rate of $0 \nu \beta \beta$ and the associated neutrino mass scale limits, depending upon the nuclear matrix element used.  

\begin{table*}[ht]
  \begin{center}
\caption{The predicted EXO-200 sensitivity to neutrinoless double beta decay, based upon nominal experimental parameters and recent nuclear matrix element calculations. \label{tab:0nsens200}}
        \begin{tabular}{|p{0.09\linewidth}|p{0.06\linewidth}|p{0.06\linewidth}|p{0.1\linewidth}|p{0.13\linewidth}|p{0.11\linewidth}|p{0.10\linewidth}|p{0.1\linewidth}|p{0.1\linewidth}|}
      \hline
          {\centering Case} & 
          {\centering Mass (ton)} & 
          {\centering Eff. (\%)} & 
          {\centering Run Time (yr)} &
          {\centering $\sigma_E/E$ @2.5 MeV (\%)} &
          {\centering Radioactive BG (events)} &
          {\centering $T_{1/2}^{0\nu}$ (yr) 90\%CL} &
          \multicolumn{2}{p{0.20\linewidth}|}{\centering Majorana Mass~(meV)\\ QRPA\cite{rodin_erratum_2007}  NSM\cite{caurier_influence_2008}} \\              
          \hline
          \hline
          EXO200 &
          0.2 &
          70 &
          2 &
          1.6 &
          40 &
          6.4$\times 10^{25}$\rule[0mm]{0mm}{3mm}&
          133 & 186 \\
          \hline
    \end{tabular}
  \end{center}
\end{table*}

The Klapdor-Kroingenthaus measurement of $0 \nu \beta \beta$ will be tested by EXO-200.  The measurement of  {$T^{0\nu}_{1/2}(Ge) = 2.23^{+0.44}_{-0.31} \times 10^{25}$ years \cite{klapdor-kleingrothaus_evidence_2006} in germanium indicates that we should see a $0\nu\beta\beta$ signal in EXO-200.  Direct comparison with this experiment requires calculated NME, however, we can estimate a range of event rates in EXO-200 by including $3\sigma$ errors on the Klapdor-Kroingenthaus measurement combined with the different NME to give best- and worst-case scenarios.  We assume here that we have 40 background events in the $0\nu\beta\beta$ region of interest for both cases and the data are taken for 2 years.  The NSM provides the most favorable NME, so using it with the lower limit on the measurement provides the upper limit on the number of $0\nu\beta\beta$ at 170 events, an 11.7 $\sigma$ signal.  The worst case is found with QRPA and the upper limit on $T_{1/2}(Ge)$, which is then 46 events for a significance of 5.0 $\sigma$.

\section{Future Possibilities}

Extensive research and development is underway for future ton-scale xenon detectors.  The EXO collaboration is examining the possibility of using xenon enriched to 80\% in $^{136}$Xe in both liquid and gas phases.  Much understanding of the technologies needed for a liquid xenon detector have come from EXO-200. In addition, a high pressure gas chamber, not low-background, is being built and will become a test bed for gas-phase detector technologies.  

One focus of the work being done is barium tagging which would allow for complete rejection of radioactive backgrounds.  The collaboration has already demonstrated a single ion trap for laser identification of barium ions \cite{flatt_linear_2007}.  We are developing a probe to attach to the barium ion in the xenon, extract the ion, and deliver it to the ion trap \cite{fierlinger_microfabricated_2008}.  Work is being done on the possibility of laser tagging the barium ion directly in gas and liquid xenon.

While the exact form of a ton-scale Xenon experiment is not yet proposed, the physics potential for such an experiment is fairly well understood.  Table \ref{tab:0nsens} shows the predicted sensitivities to $0 \nu \beta \beta$ for two different mass scales.  It is assumed that ton-scale EXO will use Barium tagging, but the sensitivities are otherwise insensitive to choice of phase used.  A limit on the effective neutrino mass at the tens to few meV scale will exclude the inverted hierarchy for the majorana neutrino.  While we will not reach those limits with EXO-200, a measurement of $T_{1/2}^{2\nu }$ will provide further information for predicting the rate of $0\nu\beta\beta$. 

\begin{table*}[ht]
     \begin{center}
\caption{Ton-Scale EXO sensitivities to neutrinoless double beta decay in two possible cases.  These assume barium-tagging removes all radioactive backgrounds but has a 70\% detection efficiency.  The aggressive estimate assumes we are successful in improving our resolution.\label{tab:0nsens}}
 \begin{tabular}{|p{0.12\linewidth}|p{0.06\linewidth}|p{0.06\linewidth}|p{0.1\linewidth}|p{0.13\linewidth}|p{0.11\linewidth}|p{0.1\linewidth}|p{0.1\linewidth}|p{0.1\linewidth}|}
         \hline
             {\centering Case} & 
             {\centering Mass (ton)} & 
             {\centering Eff. (\%)} & 
             {\centering Run Time (yr)} &
             {\centering $\sigma_E/E$ @2.5 MeV (\%)} &
             {\centering 2$\nu \beta \beta$ BG (events)} &
             {\centering $T_{1/2}^{0\nu}$ (yr) 90\%CL} &
             \multicolumn{2}{p{0.20\linewidth}|}{\centering Majorana Mass (meV)\\QRPA\cite{rodin_erratum_2007} NSM\cite{caurier_influence_2008}} \\              
             \hline
             \hline
             Conservative&
             1 &
             70 &
             5 &
             1.6&
             0.5 (use 1) &
             2$\times 10^{27}$\rule[0mm]{0mm}{3mm}&
             24 & 33 \\
             \hline
             Aggressive&
             10 &
             70 &
             10 &
             1&
             0.7 (use 1) &
             4.1$\times 10^{28}$\rule[0mm]{0mm}{3mm}&
             5.3 & 7.3 \\
             \hline
       \end{tabular}    
     \end{center}
   \end{table*}

\begin{acknowledgments}
EXO is supported by DoE and NSF in the United State, NSERC in Canada, FNS in Switzerland, and the Ministry for Science and Education of the Russian Federation in Russia.
\end{acknowledgments}

\bigskip 
\bibliography{ackerman}
\bibliographystyle{h-physrev}


\end{document}